\documentclass[letter,10pt]{article}
\usepackage{fullpage,hyperref}
\usepackage{latexsym,amsmath,amssymb, graphicx}
\usepackage{tikz}
\usetikzlibrary{shapes,arrows}

\tikzstyle{block}  =  [draw, fill=blue!20, rectangle,  minimum  height=2em,  minimum  width=4em]
\tikzstyle{sum} = [draw, fill=blue!20, circle, node distance=1cm]
\tikzstyle{input} = [coordinate] \tikzstyle{output} = [coordinate]
\tikzstyle{pinstyle} = [pin edge={to-,thin,black}]

\DeclareMathAlphabet{\mathcal}{OMS}{cmsy}{m}{n}

\newcommand{\ltwo}{\mathbb{Z}}
\newcommand{\opA}{\mathcal{A}}
\newcommand{\opB}{\mathcal{B}}
\newcommand{\opG}{\mathcal{G}}

\newcommand{\dom}{\mathcal{D}}
\newcommand{\mathK}{\mathcal{K}}
\newcommand{\opC}{\mathcal{C}}
\newcommand{\ctrlu}{u}
\newcommand{\linf}{\mathcal{L}_{\infty}}
\newcommand{\mathW}{\mathbb{W}}

\newcommand{\dwdt}{\dot{v}}
\newcommand{\citep}{\cite}
\newcommand{\staropT}{\boldsymbol{\Gamma}}
\newcommand{\mT}{\mathcal{T}}

\newtheorem{assm}{Assumption}

\newtheorem{thm}{Theorem}
\newtheorem{lem}{Lemma}

\newtheorem{defn}{Definition}
\newtheorem{problem}{Problem}
\newtheorem{remark}{Remark}

\title{\bf \Large Sub-Optimality of a Dyadic Adaptive Control Architecture}
\author{Aditya A. Paranjape \thanks{Senior Scientist, Software Systems \& Services Research Area. Email: {\small aditya.paranjape@tcs.com}} \\
{\small Tata Consultancy Services Ltd., Pune 411013, India} \\ \\
 Soon-Jo Chung \thanks{Bren Professor of Aerospace, Department of Aerospace (GALCIT). Email: {\small sjchung@caltech.edu.}}\\
{\small California Institute of Technology, Pasadena, CA 91125.} 
}
\date{}
\begin{document}
\maketitle
\thispagestyle{empty}
\pagestyle{empty}

\begin{abstract}
The dyadic adaptive control architecture evolved as a solution
to the problem of designing control laws for nonlinear systems 
with unmatched nonlinearities, disturbances and uncertainties.
A salient feature of this framework is its ability 
to work with infinite as well as finite dimensional systems, 
and with a wide range of control and adaptive laws. In this paper,
we consider the case where a control law based on the linear quadratic
regulator theory is employed for designing the control law. 
We benchmark the closed-loop system against standard linear quadratic 
control laws as well as those based on the state-dependent Riccati equation. 
We pose the problem of designing a part of the control law as a Nehari 
problem. We obtain analytical expressions for the bounds on the 
sub-optimality of the control law.
\end{abstract}

\section{Introduction}\label{sec:Intro}
In this paper, we are concerned with the control of semilinear systems of the form
$\dot{v}(t) = \opA v(t) + \opB u(t) + f(t,v),~y(t) = \opC v(t)$, 
where $v(t)$ denotes the system state, $u(t)$ is the control input, and $y(t)$ is the output. 
The underlying state space may be finite or infinite dimensional, so that the system in 
question could consist of partial and/or ordinary differential equations (PDEs and/or ODEs).
The operators $\opA$, $\opB$ and $\opC$ are the drift, control, and output operators, respectively. The forcing term $f(t,v)$ may or may not be known to the 
control designer a priori. 

In this paper, we are interested in the case wherein $f(t,v)$ is potentially nonlinear,
not entirely known, and $f(t,v) \notin {\rm range}(\opB)$. In our earlier papers
\cite{par11pde, par18dac}, we introduced a dyadic adaptive control architecture for a 
class of such systems. An extension was later proposed \cite{par16cdc} for incorporating
optimality into this dyadic adaptive control (DAC) architecture. The specific objective 
of this paper is to analyze the optimality of the architecture formally. For 
brevity, we refer to this architecture as the sub-optimal DAC (SDAC) architecture.

\subsection{Background}
The DAC architecture evolved primarily for addressing boundary control problems in
systems of partial differential equations \cite{par11pde}. In such systems, distributed
forcing terms are naturally unmatched. It offered an alternative to Lyapunov-based 
techniques \cite{hew11,krs08,meu09,Siranosian2009,win00} which are naturally suited 
mainly to well-characterized systems with a small number of degrees of freedom. 

In contrast, DAC is able to work readily with systems with an arbitrarily large number of 
degrees of freedom, while avoiding the need for a finite-dimensional approximation 
of the PDE as part of the formulation itself. At the same time, it relies on its linear terms
(which could be destabilizing) to enable its dyadic structure.

\begin{figure}[htb]
\centering
\begin{tikzpicture}[auto, node distance=2cm,>=latex']
    \node [input, name=input] {};
    \node [sum, right of=input] (sum) {};
    \node [block, right of=sum] (filter) {$H(s)$};
    \node [output, right of=filter,node distance=3cm] (virtualnode) {};
    \node [output, right of=virtualnode, node distance=2cm] (virtualnode2) {};
    \node [block, below of=filter,node distance=1.5cm] (phalf) {Particular Half};
    \node [block, below of=virtualnode, node distance=1.5cm] (sys) {Plant};
    \node [block, below of=sys, node distance=1.5cm] (hhalf) {Homogeneous Half};
    
    \draw [draw,->] (input) -- node {$r(t)$} (sum);
    \draw [->] (sum) -- node {} (filter);
    \draw [-] (filter) -- node {$u(t)$} (virtualnode);
    \draw [-] (virtualnode) -- node {} (virtualnode2);
    \draw [->] (virtualnode) -- node {} (sys);
    \draw [->] (sys) -- node {$v(t)$} (phalf);
    \draw [->] (phalf) -| node [pos=0.98] {$-$}
    	node [near end] {$\hat{y}_p(t)$} (sum);
    \draw [->] (virtualnode2) |- node{} (hhalf);
    \draw [->] (hhalf) -| node {$\hat{v}_h(t)$} (phalf);
\end{tikzpicture}
\caption{A block diagram of the DAC framework, with the subscripts {\it p} and {\it h} denoting
signals from the particular and homogeneous components. The symbols $v(t)$, $y(t)$, and $r(t)$ denote the system state, output and reference signal, respectively.}
\label{fig:lure}
\end{figure}
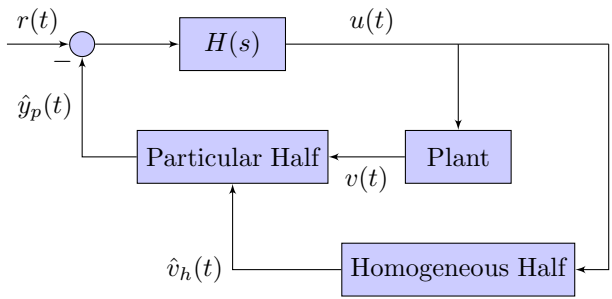

The DPO architecture, shown in Fig.~\ref{fig:lure}, uses the linear term $\opA w(t)$ 
as a pivot and decouples the system into two components, or halves. 
The {\em particular} half filters and estimates the nonlinearity, and its dynamics are not 
driven by the control signal. The {\em homogeneous} half is linear and contains the entire
control signal (the term $\opB u$) as part of its dynamics. The control law is designed 
to ensure that the output of the homogeneous half $y_h \to (r - y_p)$, where $r$ 
denotes the reference signal and $y_p$ is the output of the particular half. This is sufficient 
for ensuring that $y$ tracks $r$. The two halves are implemented in the form of observers 
which use full state feedback (i.e., $w$) to estimate the states of the two halves. The SDAC
uses, in particular, the linear quadratic regulator (LQR) theory to design the control signal
for the homogeneous half.

In this paper, we investigate the inclusion of optimality in the DAC framework. 
The theory of optimal control for linear infinite dimensional systems is well-developed
(see Chapter 6, \cite{cur95}, for instance). It has been used for solving problems such
as determining the actuator schedule in parameter-varying systems \cite{dem09} and 
for determining an optimal placement of actuators \cite{mor15}. 

Optimal control techniques rely on some knowledge of the future state of the system. This requirement, for fully-known and linear systems, is absorbed fully in the Riccati equation. 
Since this is generally not the case for uncertain, nonlinear systems, designing optimal 
controllers for such systems can be challenging. Techniques based on the state-dependent
Riccati equation \cite{mra98,cim12} have been developed to accommodate a class of
nonlinearities, but proving robustness can be difficult for such systems.
Causal approximations have been developed
for a class of systems, such as those where the reference input is known for only a part of the 
time window \cite{bar00, alb06}, or where the reference input is the output of a potentially
unknown but linear exogenous system \cite{jia12, mod14}. Techniques based on reinforcement
learning (RL), or motivated by it, have also been proposed \cite{meh09q, mod14, mod18}. 
Techniques based on RL, however, require an adequate amount of ``training'' in order to 
ensure, informally, that they stabilize the system and do not inadvertently excite the unstable
modes beyond a point. In contrast, the SDAC \cite{par16cdc} uses a dynamic system to 
generate a causal approximation.

\subsection{Contribution and Organization}
It was explained earlier in this section how the SDAC brings together LQR and  a dynamic
causal approxiation. The purpose of this paper is to investigate the sub-optimality of this
architecture systematically. 

We start by presenting the preliminaries in Sec.~\ref{sec:prelim} and the problem formulation
in Sec.~\ref{sec:pf}, including further details of DAC. We design an optimal control law
for SDAC in Sec.~\ref{sec:ocd}. In its ``pure'' form (i.e., without invoking any causal
approximations), we benchmark it against LQR (tracking) and SDRE-based
control laws. We show, in particular, that the SDAC-based law converges exponentially fast to the LQR-based law, in the 
sense of Definition~\ref{defn:claw0}. In Sec.~\ref{sec:DA}, we consider the problem of 
designing a causal approximation. We argue that it can be cast into a Nehari problem, and 
use it to provide guarantees on its sub-optimality when compared to the ``pure-form'' SDAC
law. We summarize the results of stability and robustness analysis in Sec.~\ref{sec:stab}.

\section{Preliminaries}\label{sec:prelim}
\subsection{Norms}
\begin{defn}[$\mathcal{L}_2$ norm]
We define the space $\mathcal{L}_2([0,\,T]; \mathbb{R}^n)$, where $T > 0$, as the set of functions 
$q(t) \in \mathbb{R}^n$ for $t \in [0,\,T]$ satisfying
$$
\|q\|_{\mathcal{L}_{2;T}} \triangleq \int_0^T q(t)^\top q(t)\,dt < \infty
$$
This includes the case where $T \to \infty$. The norm of $q \in \mathcal{L}_2([0,\,T]; \mathbb{R}^n)$ will be denoted
succinctly by $\|q\|_{\mathcal{L}_{2}}$ (i.e., without $T$) unless there is room for ambiguity.
\end{defn}

\begin{defn}\label{def:norm}
We define the Hilbert space $\ltwo$ with the inner product $\langle z_1,\,z_2 \rangle$ for 
$z_1,\,z_2 \in \ltwo$ and the norm $\|z\|_\ltwo = \langle z,\,z \rangle$ for $z \in \ltwo$. Corresponding to the space $\ltwo$, we define the Banach space 
$\mathW = \linf(\mathbb{R}^{+},\,\ltwo)$ with the norm
$\|v\|_{\mathW} = {\rm ess}\sup_{t \geq 0}\|v(t)\|_{\ltwo}$. We also define the truncated
norm $\|v\|_{\mathW,\tau} = {\rm ess}\sup_{0\leq t \leq \tau} \|v(t)\|_{\ltwo}$.
\end{defn}

\subsection{Operators}
\begin{defn}
The domain of an operator $\mathcal{V}$ is denoted by $\dom(\mathcal{V})$.
If $\mathcal{V}: X \to Y$ where $X$ and $Y$ are Banach spaces, 
(obviously, $\dom(\mathcal{V}) \subset X$), then we denote the induced norm 
of $\mathcal{V}$ by $\|\mathcal{V}\|_{(X,Y)}$.
If $\mathcal{V}: \mathW \to \mathW$, then we use the short-hand notation
$\|\mathcal{V}\|_i$ in place of $\|\mathcal{V}\|_{(\mathW,\mathW)}$ for ease of representation.
\end{defn}

\begin{defn}[\citep{pazbook}, Definition 1.1, Ch. 6]\label{lem:voc}
Let $\opA$ be the infinitesimal generator of the $C^0$ semigroup $\mT(t)$.
The mild solution of $\dwdt = \opA v + f(t,v)$, $v(0) = v_0 \in \ltwo$ is given by
\begin{equation}
v(t) = \mT(t)v_0 + \int_0^t \mT(t-\tau)f(\tau,v(\tau))\,d\tau,
\label{eq:ivp0mild}
\end{equation}
We succinctly denote $e^{\opA t} \triangleq \mT(t)$.
\end{defn}

\begin{defn}[Convolution]\label{def:star}
Given a $C^0$ semigroup $\mT(t)$ with the infinitesimal generator $\opA$, we define the operator 
$\staropT_\opA(t): \staropT_{\opA} (t) f(t,v(t)) = \int_0^t \mT(t-\tau) f(\tau,v(\tau))\,d\tau.$
We further define  the induced norm
$\|\staropT_{\opA}\|_i \triangleq {\rm ess}\,\sup_{(t \geq 0)} \|\staropT_A(t)\|_{i}$
\end{defn}

\begin{defn}[Inverse]\label{defn:inverse}
Let $\opA$ be the infinitesimal generator of a $C^0$ semigroup
$\mT(t)$ which we also denote as $e^{\opA t}$. The inverse of $\opA$, denoted $\opA^{-1}$,
is defined as follows: 
$$
\opA^{-1}z \triangleq -\int_0^\infty e^{\opA \tau} z\,d\tau,~~z = {\rm constant}
$$
when the right hand side exists.
\end{defn}

The proof of the next lemma is straight-forward, and omitted for brevity. 
\begin{lem}\label{lem:adjresv}
Suppose the operator $\staropT_\opA \in \mathcal{H}_\infty$; i.e., $\staropT_\opA$ maps 
$\mathcal{L}_2([0,\,T]; \ltwo) \to \mathcal{L}_2([0,\,T]; \ltwo)$
for some $T \in (0, \infty]$.
Then, the adjoint of the operator $\staropT_\opA$ in Definition~\ref{def:star} is given by
\begin{equation}
\staropT_\opA^\ast f(t) = \int_t^T e^{-\opA^\ast (t - \tau)}\,f(\tau)\,d\tau,~f(t) \in \ltwo\,\forall\,t
\end{equation}
The dependence of the adjoint on $T$ (via the definition of 
the inner product) is omitted in this sequel, unless it is absolutely necessary.
\end{lem}

\begin{defn}[Control laws]\label{defn:claw0}
A control law is a map $u: \mathbb{Z} \to \mathbb{U}$, 
and we say that two control laws $u_1$ and $u_2$ are equal 
if and only if $u_1(z) = u_2(z)$ for all $z \in \mathbb{Z}$.
\end{defn}
We distinguish a control law from a control signal $u(t)$.
The latter is a map from $\mathbb{R} \to \mathbb{U}$, obtained by 
running (or simulating) the system from a given initial 
condition. Thus, when starting from different initial 
conditions, the resulting control signals $u_1(t)$ and $u_2(t)$
need not be identical.

\begin{defn}[Further notation on operators] \label{defn:ops} 
Given a semilinear system in the standard (infinite or finite dimensional) form 
$\dwdt = \opA v + \opB u + f(t,v),~~y = \opC w$, we define the input-output 
operator
$$
\opG(\opA,\opB,\opC) u = \opC \staropT_\opA \opB u
$$
Furthermore, we define $\opG^\ast(\opA,\opB,\opC) = \opB^\ast \staropT^\ast_{\opA}\opC^\ast$.
\end{defn}

\section{Problem Formulation}\label{sec:pf}
\subsection{Plant Model}
This paper is concerned with semilinear systems of the form
\begin{eqnarray}
\nonumber && \dot{v} = \opA v + \opB u + f(v),
~v(0) = v_0 \\
&& y(t) = \opC v
\label{eq:sysp}
\end{eqnarray}
where $u \in \mathbb{R}^{n_u}$, $y \in \mathbb{R}^{n_y}$ and $v_0 \in \ltwo$, a suitably chosen Hilbert space consisting
of $\mathbb{R}^n$-valued functions.
The operators $\opB$ and $\opC$ are bounded on their respective domains. 
The operator $\opA$ is the infinitesimal 
generator of an exponential semi-group.
The control objective is to design $u(t)$ so that (i) the quadratic penalty function
$J = \int_0^T ((y - r)^\top (y-r) + u^\top R u)\,dt$, where $R > 0$, is minimized, and 
(ii) the resulting closed-loop system is stable and robust (in a sense which will be made precise later).

The minimization of the cost function does not guarantee asymptotic tracking by itself. One way to 
get around this problem is to add $\int_0^t (y - r)\,dt$ as a state in the spirit of 
the internal model principle (see \cite{cim12}, for instance). We will use a different 
approach in the paper.

\begin{assm}\label{assm:stabilizability}
The system $(\opA,\,\opB)$ is exponentially stabilizable. Moreover, 
the initial conditions are restricted to $\|v_0\|_{\ltwo} < \rho_0$ and
$v_0 \in \dom(\opA)$.
\end{assm}

\begin{assm}\label{assm:lipschitz}
The nonlinearity can be expressed as 
$$
f(v) = \alpha \phi(v)
$$
where $\phi(\cdot)$ is a $C^1$ function of $v$ and $\alpha \in \mathbb{R}^n$ is constant, but unknown 
with a known bound $\|\alpha\|_\infty < \nu_\alpha$. We also assume that its rate of change is also bounded. 

Moreover, for every $\rho > 0$, we assume that there exist $\nu_{\phi,1}(\rho),\,
\nu_{\phi,2}(\rho) \in \mathK(\rho)$ such that if $\|v(t)\|_{\ltwo} < \rho$ for some $t > 0$, then
$\|\phi(v)\|_{\ltwo} \leq \nu_{\phi,1}(\rho)\|v(t)\|_{\ltwo} + \nu_{\phi,2}(\rho)$. 
It follows that if $\|v\|_{\mathW,\tau} < \rho$ for some $\tau > 0$, then
$\|f(v)\|_{\mathW,\tau} \leq \nu_1(\rho)\|v\|_{\mathW^e,\tau} + \nu_2(\rho)$, 
for some constants $\nu_1(\rho)$ and $\nu_2(\rho)$. 
\end{assm}

We note that our control design technique is applicable readily \cite{par18dac} to more 
general linear combination of known basis functions: 
$f(v) = \sum_{i=1}^N \alpha_i(t) \phi_i(v)$. Since this ``extension'' would makes our 
presentation cumbersome without adding to our primary objective, we 
adhere to Assumption~\ref{assm:lipschitz}.

\subsection{Dyadic Adaptive Control}
Suppose that we design a control signal of the form 
$u(t) = -\mathK v(t) + u_R(t)$ for \eqref{eq:sysp}, where 
the term $u_R(t)$ is added for tracking purposes. Let us 
write $\opA_m = \opA - \opB\mathK$ (see Definition~\ref{defn:Am}
for further details), where $\opA_m$ generates an exponentially
decaying semi-group. The resulting closed-loop system can be viewed
as the sum of two sub-systems creating using $\opA_m$ as the 
pivot:
\begin{eqnarray}
\dot{v}_p &=& \opA_m v_p + f(v),~y_p = C^e v_p \label{eq:pa} \\
\dot{v}_h &=& \opA_m v_h + \opB u_R(t),~y_h = C^e v_h \label{eq:ha}
\end{eqnarray}
The two systems \eqref{eq:pa} and \eqref{eq:ha} are referred to as 
the {\em particular} and {\em homogeneous} halves, respectively. 
The states of these sub-systems can be estimated readily using 
observers, which rely on knowing the value of the actual state, $v$.
The dynamics of the observers for the two halves, with the 
observed states denoted by $\hat{v}_p$ and $\hat{v}_h$ 
respectively, are given by
\begin{eqnarray}
\hspace{-10mm}&& \dot{\hat{v}}_p \!=\! \opA_m \hat{v}_p + \hat{\alpha}(t)\phi(v),~\hat{y}_p = \opC \hat{v}_p \label{eq:ppe}\\
\hspace{-10mm}&& \dot{\hat{v}}_h \!=\! \opA_m \hat{v}_h + \opB u_R(t),~\hat{y}_h = \opC^e \hat{v}_h \label{eq:hpe}
\end{eqnarray}
with the initial conditions at $t = 0$ set to suitable values. 

The next assumption asserts the existence of a Lyapunov function corresponding to the generator $\mathcal{A}_m$, which
we need for constructing the adaptation law.
\begin{assm}\label{assm:lyap}
There exists a {\em self-adjoint coercive} operator $\mathcal{P} > 0$ and a constant $\lambda_P > 0$ such that $\forall t$,
\begin{eqnarray}
\nonumber \hspace{-7mm} & & \langle \opA_m z(t),\mathcal{P}z(t) \rangle_{\ltwo} 
\!+\! \langle \mathcal{P}z(t),\opA_m z(t) \rangle_{\ltwo} \leq -\lambda_P \!\langle\! z(t),\mathcal{P}z(t) \rangle_{\ltwo}, \\
\hspace{-7mm} & & \forall~z(t) \in \dom(\opA_m)
\label{eq:lyap}
\end{eqnarray}
\end{assm}
The predicted values $\hat{\alpha}(t)$ is found using the projection operator \cite{lavretsky2011projection}:
\begin{eqnarray}
\nonumber & & \dot{\hat\alpha}_{j}(t) = \gamma\,{\rm Proj}\left(\hat{\alpha}_{j},\,-\langle \mathcal{P}\tilde{v}(t), \phi_j(v)e_j \rangle_{\ltwo} \right),\\
&& |\hat\alpha_{j}(t)| < \nu_\alpha(1 + \epsilon)
\label{eq:pproj}
\end{eqnarray}
where $\epsilon \in \mathbb{R}^{+}$ is arbitrarily small; $\tilde{v} = \hat{v}_p + \hat{v}_h - v$;  
$\hat{\alpha}_{j}\in\mathbb{R}$ is the $j^{\rm th}$ component of 
$\hat{\alpha}$, $e_j$ denotes the $j^{\rm th}$ column of the $n \times n$ identity matrix, and $\gamma > 0$ is the adaptation gain.

The control design described in this section constitutes the 
dyadic adaptive control (DAC) architecture. It remains to determine
$\mathK$ and $u_R(t)$, for which we turn to tools from 
optimal control.

\subsection{Optimal Control Problem Formulation}
We show in Sec.~\ref{sec:stab} that the observer states $\hat{v}_p$ and $\hat{v}_h$ converge to $v_p$ and $v_h$, 
respectively. In order to design and analyse our optimal control law, we confine our analysis 
to the following system inspired by the converged observer:
\begin{eqnarray}
\nonumber \dot{v}_p &=& \opA_m v_p  + f(v) \\
\dot{v}_h &=& \opA v_h + \opB u(t) \label{eq:sysa}
\end{eqnarray}
We note the use of $\opA$ (rather than $\opA_m$) in the dynamics of $v_h$.
We formulate the control design problem as follows: design $u(t)$ 
to ensure that $y_h$ tracks $\sigma = r - y_p$ while minimizing
\begin{eqnarray*}
J \!=\! \int_0^T \big((y_h(t) \!-\! \sigma(t))^\top (y_h(t) \!-\! \sigma(t)) \! +\! u(t)^\top R u(t)\big)\,dt
\end{eqnarray*}
where $R > 0$ and $T \gg 1$. In Sec.~\ref{sec:stab}, we argue that the resulting 
closed-loop system is stable and robust. We refer to two classes of problems 
as an aid to our analysis:
\begin{enumerate}
\item Non-adaptive regulation: the reference signal $r(t) \equiv r$ (a constant) and $f(v)$ is known.
\item Complete problem: $r(t)$ as well as $f(v)$ are not identically zero.
\end{enumerate}

\section{Optimal Control Design and Analysis}\label{sec:ocd}
To facilitate the design of an optimal control law, we define an extended state space $\ltwo^f = \ltwo^e \oplus \mathbb{R}$, and 
define
\begin{eqnarray}
\dot{w}(t) \!=\! \left[\begin{matrix} \opA & 0 \\ 0 & 0 \end{matrix}\right]w(t) \!+\! \left[\begin{matrix} \opB \\ 0 \end{matrix}\right]\ctrlu(t),~
w(0) \!=\! \left[\begin{matrix}v_{h,0}\\ 1 \end{matrix}\right]
\label{eq:hcauchyf}
\end{eqnarray}
The control problem is equivalent to designing $u(t)$ to minimize
\begin{eqnarray}
\nonumber && \min_{\ctrlu} \int_0^T \left(\langle w(t),\,Q(t)w(t)\rangle + \langle \ctrlu(t),\,R\ctrlu(t)\rangle\right)\,dt \\
&& Q(t) = [\opC~~ - \sigma(t)]^{\ast}[\opC~~ - \sigma(t)]; 
\label{eq:lqrp} 
\end{eqnarray}
The control design mirrors the approach in (\cite{cur95},
Chapter 6). Ideally, we would like $T \to \infty$ in \eqref{eq:lqrp}. However, since the reference signal $\sigma(t)$ is arbitrary, the optimal cost may be infinite as $T \to \infty$. We avoid introducing a discount $e^{-\mu t}$ ($\mu > 0$) in the cost function.

The solution to \eqref{eq:lqrp} is given by
\begin{equation}
\ctrlu(t) = -R^{-1}\opB^{\ast}\left(\Pi(t) v(t) + q(t)\right)
\label{eq:ugeneral}
\end{equation}
where $\Pi(t)$ is the solution of the Riccati equation
\begin{eqnarray}
\nonumber && \frac{d}{dt}\left(z_2,\,\Pi(t)z_1\right)
= -\langle z_2,\,\Pi(t)\opA z_1 \rangle -\langle \opA z_2,\,\Pi(t) z_1 \rangle \\
&& - \langle \opC^e z_1,\,\opC^e z_2\rangle 
+ \langle \Pi(t)\opB R^{-1} \opB^{\ast}\Pi(t)z_1,\,z_2 \rangle 
\label{eq:ric}\\
\nonumber && \Pi(T) = 0,~~z_1,\,z_2 \in \dom(\opA)
\end{eqnarray}
and $q(t)$ is the mild solution of
\begin{equation}
\dot{q}(t) = -\left(\opA - \opB R^{-1}\opB^{\,{\ast}}\Pi(t)\right)^{\ast}q(t) + C^{e\ast} \sigma(t),~q(T) = 0 \label{eq:qbvp}
\end{equation}
The first part of \eqref{eq:ugeneral} is identical to the regulation problem with 
$\sigma \equiv 0$.
Therefore, we set $\Pi(t) \equiv \Pi$, the (steady state) solution to the {\em algebraic}
Riccati equation, under the assumption that $T$ is large, and write the
control signal as
\begin{eqnarray}
\hspace{-14mm} && \ctrlu(t) = -R^{-1}\opB^{\ast}\Pi v_h(t) - R^{-1}\opB^{\ast}
q(t) \label{eq:u1} \\
\hspace{-14mm} && \dot{q}(t) \!=\! -\left(\opA \!-\! \opB R^{-1}\opB^{\,{\ast}}\Pi\right)^{\ast}q(t) + C^{\ast} \sigma(t),\,q(T) \!=\! 0 \label{eq:qbvp1}
\end{eqnarray}

\begin{lem}[Theorem 5.1.5, Theorem 6.2.7, \cite{cur95}]\label{defn:Am}
The operator $\opA_m \triangleq \opA - \opB R^{-1}\opB^{\ast}\Pi$ generates an exponentially
stable semigroup, denoted by $e^{\opA_m t}$; i.e., there exist constants $M,\,\beta > 0$ such that $\|e^{\opA_m t}\|_i \leq Me^{-\beta t}$. 
Moreover, $\|\staropT_{\opA_m}\|_\infty$ is bounded, and $(sI - \opA_m)^{-1}\in \mathcal{H}_\infty$.
\end{lem}
\begin{defn}\label{defn:opgh}
We denote $\opG_m \triangleq \opG(\opA_m,\,\opB,\,\opC)$, and $\opG_m(0) = -\opB\opA_m^{-1}\opC$. Note that 
$\opG_m \in \mathcal{H}_\infty$.
\end{defn}
Since $q(T) = 0$, we get
\begin{equation}
q(t) = \int_T^t e^{-\opA_m^\ast (t-s)}\opC^\ast\sigma(s)\,ds =  -\staropT_{\opA_m}^\ast\opC^\ast\sigma
\label{eq:q0}
\end{equation}
Let us define $u_R = -R^{-1}\opB^\ast q(t)$, which serves the purpose of tracking (as against stabilization).
It follows that
\begin{equation}
u_R = R^{-1}\opG_m^\ast \sigma,~~y_h = \opG_mR^{-1}\opG_m^\ast \sigma + \opC e^{\opA_m t} v_h(0)
\label{eq:quy}
\end{equation}
where $\opG_m^\ast$ is defined (with respect to $\opG_m$) through Definition~\ref{defn:ops}.
The optimal cost can be written, with the inner product defined over $\mathcal{L}_2([0,T])$ as
\begin{equation}
J_1 = \langle y_h - \sigma, ~y_h - \sigma\rangle + \langle u, ~Ru \rangle
\label{eq:j1}
\end{equation}
where we have used the symbol $J_1$ to facilitate a comparison later in the paper. It is straight-forward
to show that
\begin{eqnarray}
\nonumber && J_1 = \langle \sigma,\,(\mathcal{P}^\ast\mathcal{P} + I - \mathcal{P})\sigma\rangle 
+ v_h(0)^\ast W_o v_h(0) 
\\ \nonumber &&+ \langle \opC e^{\opA_m t} v_h(0), (I - \mathcal{P})\sigma \rangle + \langle (I - \mathcal{P})\sigma, \opC e^{\opA_m t} v_h(0) \rangle \\
&& \mathcal{P} = \opG_mR^{-1}\opG_m^\ast 
\label{eq:j01}
\end{eqnarray}
Notice that $\sigma$ depends on $v_h$ when $f(v) \neq 0$ for some $v$.
\begin{lem}\label{lem:1}
Let $T < \infty$ denote the terminal time. Then,
the optimal cost $J_1$ is bounded if and only if $\sigma \in \mathcal{L}_2([0,T])$.
\end{lem}
{\flushleft{\em Proof}:} We have that $\|\opG_m\|_\infty < \infty$, from Lemma~\ref{defn:Am}. Moreover, 
$\|\opG_m^\ast\|_\infty = \|\opG_m\|_\infty$ (Proposition 3.15, \cite{dulbook}), as a result of which $\|\mathcal{P}\|_\infty < \infty$.
$\blacksquare$


Next, we investigate the optimality of the closed-loop system with the 
control law \eqref{eq:u1} and \eqref{eq:qbvp1} when compared to traditional linear quadratic regulators.
This entails a static causal approximation for $q(t)$, and provides the first glimpse into the optimality of the 
DAC architecture. In particular, we compare it directly with standard LQR, LQT and SDRE-based control laws. 

\begin{remark}
We note three important points in connection with the comparisons below. First, although the comparisons are
carried for relatively simple systems compared to \eqref{eq:sysa}, their value lies in benchmarking the SDAC.
Second, we are more interested in benchmarking the control law rather than the actual cost. This is because the cost
in most optimal control problems is, informally speaking, a means to an end (stability and robustness) rather
than an end in itself. Third, the difference between SDAC and other architectures arises, primarily, due to the feedback
of $y_p$ rather than $x_p$.
\end{remark}

\subsection{Comparison with LQR/LQT}
Consider the regulation problem, with $f(v) \equiv 0$ and $r \equiv 0$. In this special case,
the equation for $q(t)$ in \eqref{eq:qbvp} can be simplified further. We recall Eq.~\eqref{eq:q0}:
$$
q(t) = \int_T^t e^{-\opA_m^\ast (t-s)}\opC^\ast\sigma(s)\,ds
$$
We now let $T \to \infty$ and make the coordinate transform $\tau = s - t$. We also note
that $v(s) = e^{\opA_m (s-t)}v_p(t)$. This gives
\begin{eqnarray*}
q(t) = \left(\int_0^\infty e^{\opA^\ast_m \tau}
\opC^\ast\opC e^{\opA_m\tau} \,d\tau\right)\,v_p(t) = W_o v_p(t)
\label{eq:q0reg}
\end{eqnarray*}
where $W_o$ is the observability Grammian for the closed-loop system. Thus, it follows
that the optimal regulator is given by
\begin{eqnarray}
\nonumber u_{\rm reg}(t) &=& -\mathK\,v_h(t) - R^{-1}\opB^\ast W_o v_p(t) \\
&=& -\mathK\,v(t) + R^{-1}\opB^\ast(\Pi - W_o) v_p(t)
\label{eq:u0reg}
\end{eqnarray}
The classic LQR control law for the LTI system $\dot{v} = \opA v + \opB u$ is given by
$u_{\rm lqr} = -R^{-1}\opB^\ast \Pi\,v$, as in \eqref{eq:u1}.
This gives us the first result for the sub-optimality of the DAC.
\begin{lem}\label{lem:statreg}
Consider the system \eqref{eq:sysa} with $f(\cdot) \equiv 0$ and $r \equiv 0$. The optimal
control law, given by \eqref{eq:u1} and \eqref{eq:q0reg}, converges
exponentially fast to the classic LQR-based law. Moreover, the error between them is bounded if $\opB^\ast$ is bounded.
\end{lem} 
{\flushleft{\em Proof}}: Let $v^\star$ denote the trajectory generated by the classic LQR controller. 
We note that $u_{\rm reg}(t) - u_{\rm lqr}(t) = -\mathK(v(t) - v^\star(t)) + R^{-1}\opB^\ast (\Pi - W_o) v_p(t)$.   

It follows from the dynamics of $v_p$ in \eqref{eq:sysa}, together with Definition~\ref{defn:Am}, that $\|v_p(t)\|_\ltwo$ is 
bounded for all $t$ and $\|v_p\|_\ltwo \to 0$ exponentially fast. Thus, $u_{\rm reg}$ converges exponentially fast to $u_{\rm lqr}$, in the sense of Definition~\ref{defn:claw0}. 

It follows that the dynamics of $e = v - v^\star$ are given by $\dot{e} = \opA_m e + \opB R^{-1}\opB^\ast (\Pi - W_o) v_p$. Since $\opA_m$
is the generator of an exponentially stable semi-group, it follows that $\|e(t)\|_\ltwo$ is bounded for all $t$. Hence, $u_{\rm reg}(t) - u_{\rm lqr}(t)$
is bounded for all $t$. This completes the proof. $\blacksquare$

When $r(t) = r\neq 0$ for all $t$, following a similar approach as above, \eqref{eq:q0} can be solved to 
obtain
\begin{equation}
q(t) = W_o v_p(t) + (A^\ast_m)^{-1}\opC^\ast r,
\label{eq:q0track}
\end{equation}
where the operator $(A^\ast_m)^{-1}$ is defined as per Definition~\ref{defn:inverse}.
Akin to \eqref{eq:u0reg}, we get
\begin{eqnarray}
\hspace{-12mm}\nonumber && u_{\rm trk}(t) = -\mathK\,v_h(t) - R^{-1}\opB^\ast \big(W_o v_p(t) + (A^\ast_m)^{-1}\opC^\ast r \big)\\
\hspace{-12mm}& & = -\mathK\,v(t) \!+\! R^{-1}\opB^\ast (\Pi - W_o)\, v_p - R^{-1}\opB^\ast(A^\ast_m)^{-1}\opC^\ast r
\label{eq:u0trk}
\end{eqnarray}
The classic linear quadratic tracking control law, following \cite{mod14}, is given by
\begin{equation}
u_{lqt} = -\mathK\,v(t) - R^{-1}\opB^\ast(A^\ast_m)^{-1}\opC^\ast r
\label{eq:u0lqt}
\end{equation}
This gives us the following lemma, which generalizes Lemma~\ref{lem:statreg}. The proof
is identical to that of Lemma~\ref{lem:statreg}.
\begin{lem}\label{lem:stattrk}
Consider the system \eqref{eq:sysa} with $f(\cdot) \equiv 0$ and $r(t) = r$ (a constant) for all 
$t$. The optimal control law, given by \eqref{eq:u1} and \eqref{eq:q0reg}, converges
exponentially fast to the classic LQT-based law \eqref{eq:u0lqt}. 
Moreover, the error between them is bounded in the sense of $\linf$ if $\opB^\ast$ is bounded.
\end{lem} 

\subsection{Comparison with SDRE}
Next, we consider the problem where $f(v)$ is known and $r(t) = r$, a constant for all $t$. An SDRE-based control law can be obtained,
following the usual procedure as in \cite{mra98}. One important modification that we make to the process of deriving the SDRE
is that we do not merge $f(v)$ into $\opA v$. Rather, we introduce the state $w \equiv 1$, with the dynamics $\dot{w}= 0$ and $w(0) = 1$. 

In preparation for applying SDRE, we rewrite the dynamics of \eqref{eq:sysa}, with $v = v_p + v_h$ and with the output denoted as $y_{vw}$, as
\begin{equation}
\left[\begin{matrix} \dot{v} \\ \dot{w} \end{matrix}\right] = \left[\begin{matrix} \opA & f(v) \\ 0 & 0 \end{matrix}\right] \left[\begin{matrix} v \\ w \end{matrix}\right] + \left[\begin{matrix} \opB \\ 0 \end{matrix}\right]u,~y_{vw} = [C~-r]\left[\begin{matrix} v \\ w \end{matrix}\right]
\end{equation}
The control objective is essentially that of regulating $y_{vw}$ while minimizing a cost function of the form \eqref{eq:lqrp}.
Since the dynamics of $w$ are not stabilizable, we obtain the SDRE by introducing a small discount $e^{-\epsilon t}$ into the cost function. 
Subsequently, we allow $\epsilon \to 0$. We skip the steps, but note that the procedure is similar to \cite{mra98} and Sec.~\ref{sec:ocd}. 
The Riccati operator can be written as 
$$
\Pi = \left[\begin{matrix} \Pi_{11} & \Pi_{12} \\ \Pi_{12}^\ast & \Pi_{22}\end{matrix}\right] 
$$
of which $\Pi_{11}$ and $\Pi_{12}$ are of relevance to the present discussion. Using the shorthand notation introduced earlier, we write the 
expressions for $\Pi_{11}$ and $\Pi_{12}$ as
\begin{eqnarray}
\nonumber && \opA^\ast \Pi_{11} + \Pi_{11} \opA+ C^\ast C - \Pi_{11} \opB R^{-1} \opB^\ast \Pi_{11} = 0 \\
&& \opA_m^\ast \Pi_{12}(v) + \Pi_{11}f(v) - C^\ast r= 0
\end{eqnarray}
We get that $\Pi_{12} = (\opA_m^\ast)^{-1}(C^{\ast} r - \Pi_{11} f(v))$, which yields
\begin{eqnarray}
\nonumber u_{\rm sdre} &=& -\mathK v - R^{-1}B^\ast(\opA_m^\ast)^{-1}C^\ast r \\
 & & + R^{-1}\opB^\ast (\opA_m^\ast)^{-1}\Pi_{11} f(v) \label{eq:usdre1}
\end{eqnarray}
Notice that the first two terms are identical to $u_{\rm lqt}$ from \eqref{eq:u0lqt}, and the gain $\mathK$ is also identical
to the optimal controller in \eqref{eq:u1}. To facilitate a comparison, as before, we define
$$
u_{R,sdre} \!=\!  -R^{-1}B^\ast(\opA_m^\ast)^{-1}C^\ast r + R^{-1}\opB^\ast (\opA_m^\ast)^{-1}\Pi_{11} f(v)
$$ 
We recall the corresponding expression for $u_R$ from \eqref{eq:quy}:
\begin{equation}
u_R = R^{-1}\opB^\ast\staropT_{\opA_m}^\ast\opC^\ast r 
 - R^{-1}\opB^\ast\staropT_{\opA_m}^\ast \opC^\ast \opC \staropT_{\opA_m} f(v)
\label{eq:quy1}
\end{equation}
where we have assumed that $v_p(0) = 0$. Thus, we get
\begin{eqnarray}
\hspace{-7mm}&& \nonumber u_R - u_{R,sdre} = R^{-1}\opB^\ast(\staropT_{\opA_m}^\ast + (\opA_m^\ast)^{-1})\opC^\ast r 
\\
\hspace{-7mm} &&- R^{-1}\opB^\ast (\staropT_{\opA^\ast_m}\opC^\ast \opC \staropT_{\opA_m}f(v) +  (\opA_m^\ast)^{-1}\Pi_{11}f(v))
\end{eqnarray}
Since we have assumed that $r$ is constant, it can be shown that
\begin{eqnarray*}
(u_R - u_{R,sdre})_{t \to \infty} &=& -R^{-1}\opB^\ast (\staropT_{\opA^\ast_m}\opC^\ast \opC \staropT_{\opA_m}f(v) \\
&& + (\opA_m^\ast)^{-1}\Pi_{11}f(v))
\end{eqnarray*}
Here, we have compared the two control laws rather than the time histories of the two 
control signals. It is clear that, unlike the LQR/LQT case, the two control laws 
need not converge to each other.

The static approximation works well when the dynamics of the particular half
and the reference signal are known {\it a priori}: this completely mitigates the anti-causal 
nature of the boundary value problem \eqref{eq:qbvp1}. When these are unknown, the 
typical course of action (in the context of optimal control) has been to use some form of 
dynamic approximation \cite{mod14,mod18}. We opt for an alternate approach, which seeks 
to approximate the backward-in-time equation dynamics of $q$ in \eqref{eq:u1} by a 
forward-in-time approximation.

\section{Dynamic Causal Approximation for Adaptive Systems}\label{sec:DA}
It is a well-known property of the solution to the LQR problem that the adjoint state evolves on the 
stable manifold of the combined system-adjoint dynamics, and the stable eigenvalues are precisely 
those of $\mathcal{A}_m$. This motivates us to express the control signal using the 
following {\em finite-dimensional} dynamic approximation:
\begin{eqnarray}
\ctrlu(t) &=& -R^{-1}\opB^{\ast}\Pi v_h(t) + R^{-1}H_C p(t), \label{eq:da1} \\
\dot{p}(t) &=& H_A p(t) + H_B \sigma(t),~p(0) = p_0 \in \mathbb{R}^{n_p} \nonumber
\end{eqnarray}
where $H_A \in \mathbb{R}^{n_p\times n_p}$ is Hurwitz, and $H_B,\,H_C^{\top} \in \mathbb{R}^{n_p}$. We set $p(0) = 0$, which is in contrast to the 
approach taken in other causal approximations, such as \cite{bar00}.

We define
\begin{equation}
\opG_H = H_C\staropT_{H_A}H_B 
\label{eq:opgh}
\end{equation}
We state the dynamic approximation problem as follows.
\begin{problem} \label{prob:DA1}
Determine the system ($H_A,\,H_B,\,H_C$; $H_A \in \mathcal{H}_\infty$) which minimizes the cost function and ensures that the closed-loop system in stable.
\end{problem}

From the dynamics of $p$ in \eqref{eq:da1}, with $p(0) = 0$, we get
\begin{equation}
u_R(t) = R^{-1}\opG_H \sigma(t),~~y_h = \opG_m R^{-1} \opG_H\, \sigma(t)
+ \opC e^{\opA_m t} v_h(0) \label{eq:mildp1}
\end{equation}
A comparison with the expression for $u_R$ from \eqref{eq:quy} suggests that we compute a dynamic approximation by solving the Nehari problem:
\begin{equation}
\opG_H^{N} = {\rm arg}\,\min_{X \in \mathcal{H}_{\infty}} \| \opG_m^{\ast} - X\|_\infty
\label{eq:nehari01}
\end{equation}
This allows us to bound the error in the cost function with the dynamic approximation when compared to $J_1$ in \eqref{eq:j1}.
Let $J_2$ denote the cost function accumulated by employing the controller in \eqref{eq:da1}. Then, we have that
\begin{equation}
J_2 = \|(\opG_mR^{-1}\opG_H^N - I_{n_y})\sigma\|_{\mathcal{L}_2} + \|R^{1/2}u\|_{\mathcal{L}_2}
\label{eq:j2}
\end{equation}
where 
\begin{eqnarray}
\nonumber u &=&  u_R - \mathK v_h \\
&=& (I_{n_u} - \mathK \staropT_{\opA_m}\opB)R^{-1}\opG_H^N\sigma - 
\mathK e^{\opA_m t} v_h(0)
\label{eq:uda2}
\end{eqnarray}

Note that $\sigma$ depends on $v_p$ and hence on $v_h$ (through the nonlinear function
$f(v)$.). This can complicate the comparison of $J_2$ and $J_1$ substantially. A simple 
comparison can be obtained for the case where $f(v)$ is actually independent of $v$; i.e., 
where it is a pure exogenous disturbance.
\begin{thm}\label{thm:diffopt}
Let $\sigma(t) \in \mathcal{L}_2([0,T])$ be an exogenous signal and independent of the system state $v_h$, and $T > 0$. Consider the cost functions $J_1$ and $J_2$, evaluated in \eqref{eq:j1} and \eqref{eq:j2} with $v_h(0)$ being identical in both
cases. We have that $|J_2 - J_1| \leq k\|\sigma\|_{\mathcal{L}_2}$, where
$k < \infty$.
\end{thm}
{\flushleft{\em Proof}}: We recall the triangle inequality: $\|z_1\| - \|z_2\| \leq \|z_1 - z_2\|$.
We apply it to each of the two terms in the cost functions $J_1$ and $J_2$. This gives
\begin{eqnarray}
\nonumber |J_2 - J_1| &\leq& \|(\opG_mR^{-1}\opG_H^N - \opG_mR^{-1}\opG_m^{\ast})\sigma\| \\
&& + \|R^{-1/2}(\opG_m^\ast - \opG_H^N)\sigma\|,
\end{eqnarray}
from which it follows that 
\begin{eqnarray}
|J_2 - J_1| &\leq& S\|\sigma\|_{\mathcal{L}_2} \label{eq:costdiff} \\
\nonumber S &=& (\|\opG_m\|_\infty + \|R^{-1/2}\|_\infty)\|\boldsymbol{\theta}_{\opG_m^\ast}\|
\end{eqnarray}
where $\boldsymbol{\theta}_{\opG_m^\ast}$ is the Hankel operator for $\opG_m$.
This completes the proof.
$\blacksquare$

Theorem~\ref{thm:diffopt} calculates an explicit formula for an upper bound on the sub-optimality induced by the 
dynamic causal approximation, in comparison to the ``pure form'' control law of \eqref{eq:quy}. 

\begin{remark}
In Theorem~\ref{thm:diffopt}, the effect of $v$ on $\sigma$ is implicitly ignored. In order to factor in the effect of 
$v$, it is essential to prove that the closed-loop system is stable. This is the subject of the next section. We use 
stability in the sense of $\linf$ rather than $\mathcal{L}_2$.
\end{remark}

\begin{remark}[Asymptotic tracking] 
Although we minimize the cost of tracking, it does not guarantee asymptotic tracking even when the reference input is a 
constant. In order to ensure asymptotic tracking, one can either extend the state space to include the tracking error 
$y_h - \sigma$ as a state (as in \cite{cim12}), or one can solve the constrained Nehari problem
\begin{equation}
\opG_H^{N} = {{\rm arg}\,\min}_{X \in \mathcal{H}_{\infty}; \opG_m(0)R^{-1}\hat{X}(0) = I_{n_y}} \| \opG_m^{\ast} - X\|_\infty
\label{eq:nehari02}
\end{equation}
where $\hat{X}$ denotes the Laplace transform of $X$. This option, of course, worsens the bound in Theorem~\ref{thm:diffopt}.
\end{remark}

\section{Analysis of the Complete Closed-Loop System}\label{sec:stab}
\subsection{Summary of the Closed-Loop System}\label{sec:ctrlsummary}
The closed-loop system consists of the original system \eqref{eq:sysp} and the controller. 
If dynamic causal approximation is employed, the controller consists of the primary control law
\begin{eqnarray} 
\ctrlu(t) &=& -R^{-1}\opB^{\ast}\Pi v(t) + R^{-1}H_Cp(t)
\label{eq:uptsum} \\
\dot{p}(t) &=&H_A p(t) + H_B (r(t) - \hat{y}_p(t)),~p(0) = p_0 
\label{eq:pbvpsum}
\end{eqnarray}
and the observers equipped with the projection operator \cite{lavretsky2011projection}:
\begin{eqnarray*}
\hspace{-10mm}&& \dot{\hat{v}}_p \!=\! \opA_m \hat{v}_p + \hat{\alpha}(t)\phi(v),~\hat{y}_p = \opC \hat{v}_p \label{eq:ppesum}\\
\hspace{-10mm}&& \dot{\hat{v}}_h \!=\! \opA_m \hat{v}_h - \opB R^{-1} H_cp(t),~\hat{y}_h = \opC \hat{v}_h \label{eq:hpesum} \\
\nonumber & & \dot{\hat\alpha}_{j}(t) = \gamma\,{\rm Proj}\left(\hat{\alpha}_{j},\,-\langle \mathcal{P}\tilde{v}(t), \phi_j(v)e_j \rangle_{\ltwo} \right),\\
&& |\hat\alpha_{j}(t)| < \nu_\alpha(1 + \epsilon)
\label{eq:pprojsum}
\end{eqnarray*}

\subsection{Stability Analysis}
We recall the following results from \cite{par16cdc} and \cite{par18dac} for completeness. These results
prove the stability of the closed-loop system subject to a small gain condition.

\begin{lem}\label{lem:obserr}
Suppose that $\|v\|_{\mathW,t} < \rho_w$ for some constant $\rho_w > 0$. Then,
the observation error $\|\tilde{v}(t)\|_{\ltwo}$ is uniformly bounded and 
the bound can be made arbitrarily small by increasing $\gamma$.
Furthermore, the observation errors $\|\tilde{v}_p(t)\|_\ltwo$ and $\|\tilde{v}_h\|_\ltwo$ are uniformly bounded,
and can be made arbitrarily small by increasing $\gamma$.
\end{lem}

Next, we assert that the control input $\ctrlu(t)$ is bounded. Recall that the second term in the control signal \eqref{eq:uptsum} by $\ctrlu_R(t) = H_C p(t)$. 
Let $H(s) = H_C(sI-H_A)H_B$ denote the transfer function between $U_Rr(s)$ and $(R(s) - \hat{Y}_p(s))$.
\begin{lem}\label{lem:u}
Let $\|v\|_{\mathW, t} < \rho_w$ for some $t$ and $\rho_w > 0$. Then,
the control input $\ctrlu(t)$ is bounded and a $C^1$ function of time. Moreover, 
there exist constants 
$\delta_{iw} \equiv \delta_{iw}(H(s),\rho)$,  $\delta_{ir}\equiv \delta_{ir}(H(s),\rho)$ 
and $\delta_{iu}\equiv \delta_{iu}(H(s),\rho)$ for $i = 0,\,1$ such that
$\|u_R\|_{\linf,\tau} \leq \delta_{0w}\|v\|_{\mathW,\tau} + \delta_{0r}\|r\|_{\linf,\tau} + \delta_{0u}$. 
\end{lem}

Next, we define a small gain condition.
\begin{assm}[Small-gain condition] \label{assm:smallgain} We assume that there exists a constant $\rho_w$, an arbitrarily small $\epsilon_s > 0$, and a stable strictly proper 
$H(s)$ such that the following inequality is satisfied:
$$
\frac{M\rho_0 + \|\staropT_{\opA_m}\|_i(\nu_2(\rho_w) \!+\! \delta_{0r}\|r\|_{\linf} \!+\! \delta_{0u})}{1 - \|\staropT_{\opA_m}\|_i(\nu_1(\rho_w) + \|\opB\|_{i}(\delta_{0w})}\leq \rho - \epsilon_s
$$
where the constants have been defined in Assumption~\ref{assm:lipschitz} and Lemma~\ref{lem:u}.
\end{assm}

Finally, we state the main result of this section. 
\begin{thm}\label{thm:main1}
The closed-loop system summarized in Sec.~\ref{sec:ctrlsummary} is well-posed and
bounded-input-bounded-state stable in the sense of $\linf$ if Assumption~\ref{assm:smallgain} is satisfied.
\end{thm}
Notice that the satisfaction of the small gain condition also endows the closed-loop system
with robustness. The proof of well-posedness relies on Theorems 6.1.4 and 6.1.5 from \citep{pazbook}.

\subsection{Bounds on the Cost}
We would like to derive some bounds on the cost function. In a nonlinear setting, the baseline cost is $J_1$ from the pure form
solution of Sec.~\ref{sec:ocd}. Lemma~\ref{lem:obserr} leads us to conclude that there exists a constant $\rho_{\rm obs}$ 
which can be made arbitrarily small, such that
$$
\|\hat{y} - y\|_{\linf,T} \leq \rho_{\rm obs}
$$
Thus, the cost calculated using Theorem~\ref{thm:diffopt} and \eqref{eq:j1} is a reasonable estimate of the cost incurred by 
the closed-loop system. We note that an exact expression can be found when $\phi(v) = 1$. When $f(v) = \mathcal{S}v$
for some linear $\mathcal{S}$, expressions similar to \eqref{eq:quy} and \eqref{eq:j1} would feature a semigroup 
generated by the time-varying operator $\opA_m + \hat{\alpha}(t)\mathcal{S}$. Even in that case, since the same 
semigroup would feature in the computation of $J_2$, the expression for the difference would be similar to 
\eqref{eq:costdiff}.

\section{Conclusion}
We analyzed the optimality of LQR-based tracking control laws 
designed for semilinear systems in the dyadic adaptive control
(DAC) framework. We showed that a ``pure form''
law (i.e., without any dedicated causal approximation)
converges exponentially to standard LQR laws for LTI systems.
We determined analytical expressions for the cost function
in the presence of a dynamic compensator for causal 
approximation for a class of perturbed linear systems. The
compensator design problem was posed, in particular, in the 
framework of model-matching problems. While an exact expression
for the cost function in the presence of nonlinear forcing 
remains elusive, the present paper has amply demonstrated
the utility of the dyadic adaptive architecture for accommodating
optimality, robustness and adaptation systematically in a 
single framework.

\bibliography{Bibliography}

\begin{thebibliography}{10}

\bibitem{alb06}
R.~{Alba-Flores} and E.~Barbieri.
\newblock Real-time infinite horizon {Linear-Quadratic} tracking controller for
  vibration quenching in flexible beams.
\newblock In {\em Proc. {IEEE} Conference on Systems, Man, and Cybernetics,
  Taipei, Taiwan}, pages 38 -- 43, 2006.

\bibitem{bar00}
E.~Barbieri and R.~{Alba-Flores}.
\newblock On the infinite-horizon {LQ} tracker.
\newblock {\em Systems and Control Letters}, 40:77 -- 82, 2000.

\bibitem{cim12}
T.~Cimen.
\newblock Survey of state-dependent {Riccati} equation in nonlinear optimal
  feedback control synthesis.
\newblock {\em Journal of Guidance, Control, and Dynamics}, 35(4):1025 -- 1047,
  2012.

\bibitem{cur95}
R.~F. Curtain and H.~J. Zwart.
\newblock {\em An Introduction to Infinite-Dimensional Linear Systems Theory}.
\newblock Texts in Applied Mathematics (Vol. 21). Springer-Verlag, 1995.

\bibitem{dulbook}
G.~E. Dullerud and F.~Paganini.
\newblock {\em A Course in Robust Control Theory: A Convex Approach}.
\newblock Springer, 2000.

\bibitem{hew11}
W.~He, S.~S. Ge, B.~V.~E. How, Y.~S. Choo, and K.~S. Hong.
\newblock Robust adaptive boundary control of a flexible marine riser with
  vessel dynamics.
\newblock {\em Automatica}, 47:722 -- 732, 2011.

\bibitem{dem09}
O.~V. Iftime and M.~A. Demetriou.
\newblock Optimal control of switched distributed parameter systems with
  spatially scheduled actuators.
\newblock {\em Automatica}, 45(2):312 -- 323, 2009.

\bibitem{jia12}
Y.~Jiang and {Z-P} Jiang.
\newblock Computational adaptive optimal control for continuous-time linear
  systems with completely unknown dynamics.
\newblock {\em Automatica}, 48:2699 -- 2704, 2012.

\bibitem{mod18}
B.~Kiumarsi, K.~G. Vamvoudakis, H.~Modares, and F.~L. Lewis.
\newblock Optimal and autonomous control using reinforcement learning: A
  survey.
\newblock {\em {IEEE} Transactions on Neural Networks and Learning Systems},
  29(6):2042 -- 2062, 2017.

\bibitem{krs08}
M.~Krstic and A.~Smyshlyaev.
\newblock {\em Boundary Control of {PDE}s: A Course on Backstepping Designs}.
\newblock Advances in Design and Control, {SIAM}, 2008.

\bibitem{lavretsky2011projection}
E.~Lavretsky, T.~E. Gibson, and A.~M. Annaswamy.
\newblock Projection operator in adaptive systems, 2011.
\newblock arXiv preprint arXiv:1112.4232.

\bibitem{meh09q}
P.~G. Mehta and S.~Meyn.
\newblock {Q-Learning} and {Pontryagin's} minimum principle.
\newblock In {\em Proceedings of the $48^{\rm th}$ {IEEE} Conference on
  Decision and Control (CDC)}, pages 3598 -- 3605, 2009.

\bibitem{meu09}
T.~Meurer and A.~Kugi.
\newblock Tracking control for boundary controlled parabolic pdes with varying
  parameters: Combining backstepping and differential flatness.
\newblock {\em Automatica}, 45:1182 -- 1194, 2009.

\bibitem{mod14}
H.~Modares and F.~L. Lewis.
\newblock Linear quadratic tracking control of partially-unknown
  continuous-time systems using reinforcement learning.
\newblock {\em {IEEE} Transactions on Automatic Control}, 59(11):3051 -- 3056,
  2014.

\bibitem{mor15}
K.~Morris, M.~A. Demetriou, and S.~D. Yang.
\newblock Using {$H_2$}-control performance metrics for the optimal actuator
  location of distributed parameter systems.
\newblock {\em IEEE Transactions on Automatic Control}, 60(2):450 -- 462, 2015.

\bibitem{mra98}
C.~P. Mracek and J.~R. Cloutier.
\newblock Control designs for the nonlinear benchmark problem via the
  state-dependent {Riccati} equation method.
\newblock {\em International Journal of Robust and Nonlinear Control}, 8:401 --
  433, 1998.

\bibitem{par16cdc}
A.~A. Paranjape and {S.-J.} Chung.
\newblock Sub-optimal boundary control of semilinear {PDEs} using a dyadic
  perturbation observer.
\newblock In {\em Proc. $55^{\rm th}$ {IEEE} Conference on Decision and Control
  ({CDC}), Las Vegas, NV}, pages 1382 -- 1387, 2016.

\bibitem{par18dac}
A.~A. Paranjape and {S.-J.} Chung.
\newblock Robust adaptive boundary control of semilinear {PDE} systems using a
  dyadic controller.
\newblock {\em International Journal of Robust and Nonlinear Control},
  28(8):3174 -- 3188, 2018.

\bibitem{par11pde}
A.~A. Paranjape, J.~Guan, {S-J.} Chung, and M.~Krstic.
\newblock {PDE} boundary control for flexible articulated wings on a robotic
  aircraft.
\newblock {\em {IEEE} Transactions on Robotics}, 29(3):625 -- 640, 2013.

\bibitem{pazbook}
A.~Pazy.
\newblock {\em Semigroups of Linear Operators and Applications to Partial
  Differential Equations}.
\newblock Applied Mathematical Sciences; v.44. {Springer-Verlag}, New York,
  1983.

\bibitem{Siranosian2009}
A.~A. Siranosian, M.~Krstic, A.~Smyshlyaev, and M.~Bememt.
\newblock Motion planning and tracking for tip displacement and deflection
  angle for flexible beams.
\newblock {\em Journal of Dynamic Systems, Measurement, and Control},
  131(031009), 2009.

\bibitem{win00}
J.~J. Winkin, D.~Dochain, and P.~Ligarius.
\newblock Dynamical analysis of distributed parameter tubular reactors.
\newblock {\em Automatica}, 36:349 -- 361, 2000.

\end{thebibliography}
\bibliographystyle{plain}

\end{document}